\documentclass[USenglish,twocolumn]{article}

\usepackage[utf8]{inputenc}
\usepackage[big]{dgruyter}
\usepackage{booktabs}
\usepackage{float}

\newcommand\numbereq{\addtocounter{equation}{1}\tag{\theequation}}

\begin{document}

  \articletype{Research Article{\hfill}Open Access}

  \author*[1]{Zackary L. Hutchens}

\author[2]{Brad N. Barlow}
\author[3]{Alan Vasquez Soto}
\author[4]{Dan E. Reichart}
\author[6]{Josh B. Haislip}
\author[7]{Vladimir V. Kouprianov}
\author[8]{Tyler R. Linder}
\author[9]{Justin P. Moore}

  \affil[1]{High Point University, E-mail: zhutchen@highpoint.edu}
  \affil[2]{High Point University, E-mail: bbarlow@highpoint.edu}
  \affil[3]{High Point University, E-mail: vasqua14@highpoint.edu} 
  \affil[4]{University of North Carolina at Chapel Hill, E-mail: dan.reichart@gmail.com}

  \title{\huge New Pulse Timing Measurements of the sdBV Star CS 1246}

  \runningtitle{New Pulse Timing Measurements of the sdBV Star CS 1246}


  \begin{abstract}
{CS 1246 is a hot subdwarf B star discovered in 2009 to exhibit a single, large-amplitude radial pulsation. An O-C diagram constructed from this mode revealed reflex motion due to the presence of a low--mass M dwarf, as well as a long--term trend consistent with a decrease in the pulsational period. The orbital reflex motion was later confirmed with radial velocity measurements. Using eight years of data collected with the Skynet Robotic Telescope Network, we show that the pulsation amplitude of CS 1246 is decaying nonlinearly. We also present an updated O-C diagram, which might now indicate a positive $\dot P$ and a new $2.09 \pm 0.05$ yr oscillation consistent with orbital reflex motion of the entire inner sdB+dM binary, possibly due to the gravitational influence of a circumbinary planet with minimum mass $m\sin i = 3.3 \, \pm \, ^{2.1}_{1.2}$ $M_{\rm Jup}$. However, unlike the presence of the M dwarf, we hesistate to claim this object as a definitive detection since intrinsic variability of the pulsation phase could theoretically produce a similar effect.}
\end{abstract}


  \journalname{Open Astronomy}
\DOI{DOI}
  \startpage{1}
  \received{Sep 30, 2017}
  \revised{Nov 13, 2017}
  \accepted{Nov 16, 2017}

  \journalyear{2017}
  \journalvolume{1}

\maketitle

\section{Introduction}

Hot subdwarf B (sdB) stars are evolved, helium-burning objects located on the extreme horizontal branch \citep{Heber2016}. Although some sdBs might form via white dwarf mergers, most are thought to be the remnant cores of red giants whose outer hydrogen envelopes were removed by external processes. Binary star interactions including Roche lobe overflow (RLOF) and common envelope (CE) evoluton are the  likely culprits \citep{Han2002, Han2003}. Models predict that sdBs will evolve directly into white dwarfs upon core helium depletion.

The first rapidly--pulsating hot subdwarfs (sdBV$\rm _r$) were discovered more than twenty years ago \cite{Kilkenny1997} and have since provided a means to probe the internal structure and physical interactions of sdB stars. Specifically, sdB p--mode pulsations are stable enough to be used as precise clocks to measure small evolutionary changes in the pulsation frequencies, or to detect reflex motion due to orbiting companions. Planets as small as 1.9$M_{\rm Jup}$ have been reported using observed--minus--calculated (O-C) diagram analyses of sdBV$\rm _r$ pulsation modes \cite{Silvotti2007,Baran2015}.

CS 1246 is a hot subdwarf B star which was discovered to pulsate in 2009 through both multicolor photometric and time--resolved spectroscopic observations \citep{barlow2010}. Model fits to the average spectrum revealed atmopsheric values of $\log g = 5.46\pm 0.11$ and $T_{\rm eff} = 28450 \pm 700$ K. The Fourier transform of the light curve showed a single, large-amplitude pulsation mode, likely $\ell = 0$, with a period of $371.707 \pm 0.002$ s. Application of the Baade--Wesselink method led to mass and stellar radius estimations of $M = 0.39\pm \, ^{0.30}_{0.13}M_\odot$ and $R=0.19\pm 0.08R_\odot$ respectively. 

Due to its single pulsation mode, CS 1246 was a prime candidate for a follow-up pulse-timing analysis. An O-C diagram constructed in 2011 from two years of data showed a downward parabolic trend consistent with a period decrease of $\dot P = -1.9 \times 10^{-11}$ s s$^{-1}$, along with a 14.1 d sinusoidal variation \citep{barlow2011}. The latter suggested the presence of an orbiting companion, probably an M dwarf, which was later confirmed with radial velocity measurements from SOAR/Goodman \cite{barlow2011_RV}.

In this paper, we present six additional years of pulse timing measurements, obtained with the Skynet Robotic Telescope Network, and construct an updated O-C diagram. We find that the pulsation amplitude of CS 1246 has been decaying nonlinearly since its discovery in 2009. Additionally, we find that the pulsational period might be {\em increasing} and present evidence in favor of a third, low--mass circumbinary object in the system.

\section{Observations}

Up to the present, we have acquired 238 total light curves of CS 1246 spanning more than eight years, from April 2009 to July 2017; the first 89 light curves were published in 2011 \cite{barlow2011}. The vast majority of our data are white--light observations carried out with the Panchromatic Robotic Optical Monitoring and Polarimetry Telescopes (PROMPT) with SKYNET, an online, fully-automated scheduling software through which observing requests are prioritized in a queue. Detailed information regarding PROMPT and SKYNET can be found in Reichart et al. 2005 \cite{Reichart2005}. Twelve of the PROMPT light curves represent simultaneous, multi--color photometry obtained over three consecutive nights in April 2009 using Sloan $u'$, $g'$, $r'$, and $i'$ filters, for the purposes of mode identification \citep{barlow2010}. Three additional observations were obtained with the Goodman spectrograph on the 4.1-m Southern Astrophysical Research (SOAR) telescope \cite{Clemens2004} using a Johnson $V$ filter. Most light curves were a few hours in length with average integration and cycle times of 30 s and 36 s, respectively.

All images were bias--subtracted and flat--fielded using standard procedures in {\sc IRAF}\footnote{IRAF is distributed by the National Optical Astronomy Observatories, which are operated by the Association of Universities for Research in Astronomy, Inc., under cooperative agreement with the National Science Foundation.}. We then extracted aperture photometry by selecting aperture radii that optimized signal-to-noise (S/N) ratios in each light curve. Counts from the sky were subtracted using sky annuli, and transparency variations were removed using the light curves of nearby, constant comparison stars. We also accounted for residual atmospheric extinction effects by fitting and normalizing the light curves with parabolas. All times were converted to Barycentric Dynamical Time (BJD$_{\rm TDB}$) using Jason Eastman's online calculator \cite{Eastman2010}.

\section{Amplitude Decay}

\begin{figure}
\includegraphics[scale=.6]{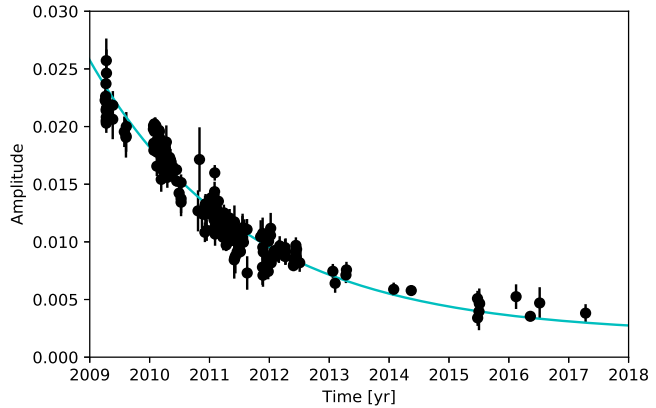}
\caption{Amplitudes of the 371.7 s pulsation mode taken from least--squares fits to individual light curves from PROMPT. The solid line denotes the best--fitting exponential function to the data, illustrating a clear, nonlinear decay over an eight--year period. Only amplitudes taken from ``white light'' measurements through clear or open filters are included. \label{fig1}}
\end{figure}
 We used the Levenberg-Marquardt algorithm \cite{lm1963} with weighted uncertainties to fit sine waves to our CS 1246 light curves, extracting a pulsation amplitude for each observation. An exponential decay with mean lifetime $2.6 \pm 0.1$ yr, shown in Figure 1, emerges when these amplitudes are plotted over time. We note that only white--light measurements are used in this analysis  since sdBV$_{\mathrm r}$ pulsation amplitudes are wavelength--dependent. While CS 1246 was once a strong pulsator, with an amplitude around 24 ppt, our most recent measurement reveals an amplitude just above 4 ppt. The pulsations are now barely visible above the noise level in an FT of a single night's light curve from PROMPT. 
    
\begin{figure*}[t]
\begin{center}
\includegraphics[scale=.65]{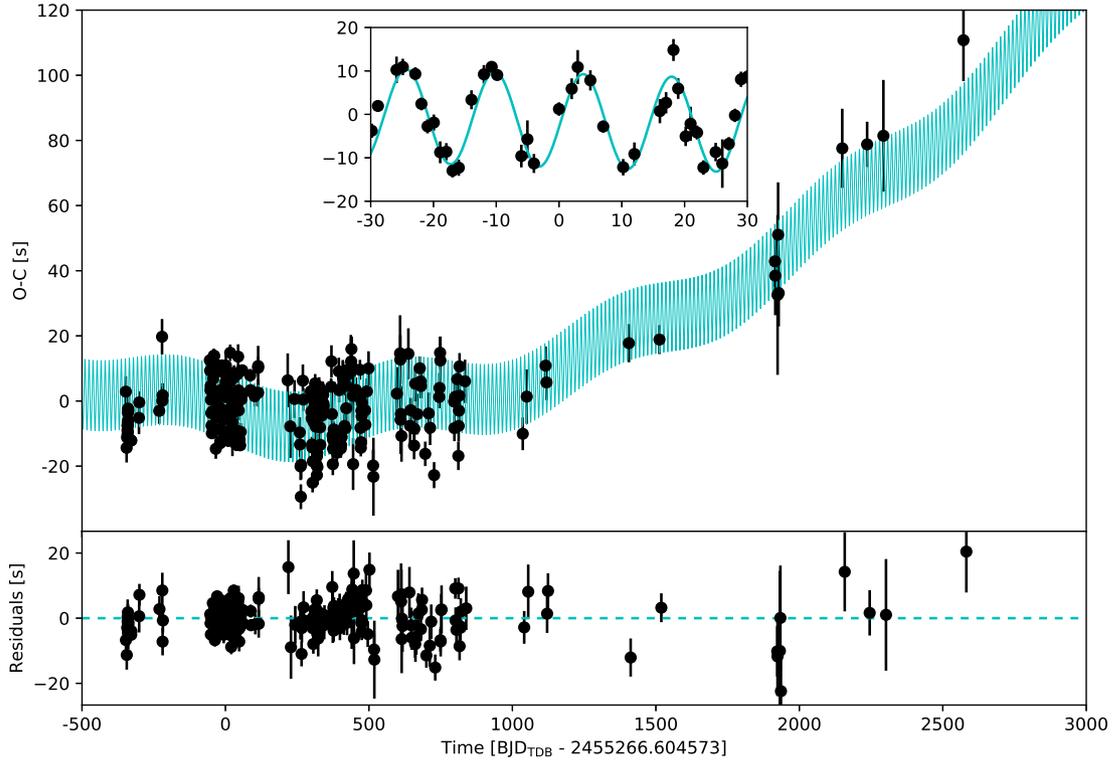}
\caption{The O-C diagram of CS 1246, constructed from its 371.7 s pulsation mode. Three major patterns dominate the diagram: (i) a concave-up parabola, indicating a period increase, (ii) a short--period sine wave from reflex motion due to the previously--confirmed M dwarf companion, and (iii) a long-period sine wave, consistent with reflex motion of the entire inner sdB+dM binary, possibly due to a circumbinary planet. The solid lines denotes our best--fitting model to the data (Equation 2). The inset highlights the strong 14.1-d oscillation from the M dwarf companion. The bottom panel shows residuals of the data and the nonlinear fit (Eq. 2). }
\end{center}
\end{figure*}    
    
    Previous studies show that sdBV$\rm _r$ stars commonly exhibit amplitude variations \cite{Kilkenny2010}, but variations are typically observed in pulsators with {\em multiple} oscillation modes, between which energy can be transferred. No other obvious pulsation modes are evident in our most recent CS 1246 data; as far as we can tell from ground observations, it remains a single--mode pulsator. However, as most light curves obtained are only 1--3 hours in length and were individually normalized with parabolas for this study, we do not have the ability to comment on the presence of hydrid g--mode pulsations in CS 1246. Careful corrections for differential atmospheric extinction are necessary before we weld light curves together to look for low--amplitude, low--frequency pulsations, which we might attempt in the future. We note that TESS, with its 2--min sampling, may be able to shed light on the potential hybrid nature of CS 1246. 



\section{The O--C Diagram}




The single pulsation mode of CS 1246 makes it a strong candidate for O-C diagram analyses. We used a nonlinear, least-squares fitting algorithm \cite{lm1963} to fit sine waves to all 238 light curves in order to obtain phase measurements. We note that phase differences can arise when comparing brightness variations in different wavebands, especially at extremely blue wavelengths. For radial pulsations, however, theory predicts the most extreme phase shift in the $u'$ band to be only a couple of seconds \citep{randall2005}, much less than the $\sim$8 s uncertainties associated with those particular measurements. As such, we use all data in our timing analysis. 

\begin{table*}[t]
\centering
\begin{tabular}{c c c c l}
 Parameter & Value & Error & Unit & Comment\\ \hline
 $T_0$ & $2455266.604573$ & $\pm 0.000005$ & BJD$\rm _ {TDB}$ & reference time of light maximum\\
 $P$ & $371.691831$ & $\pm 0.000006$ & s & fundamental pulsation period at $T_0$\\
 $\dot P$ & $1.74 \times 10^{-12}$ & $\pm 0.08 \times 10^{-12}$ & s s$^{-1}$ & pulsation period change\\
 $A$ & $10.7$ & $\pm 0.5$ & s & fortnightly phase-variation semi-amplitude\\
 $\Pi$ & $14.103$ & $\pm 0.003$ & d & fortnightly phase-variation period\\
 $\phi$ & $-0.15$ & $\pm .06$ & rad & fortnightly phase-variation phase\\
 $B$ & $4.1$ & $\pm 0.7$ & s & biennial phase-variation semi-amplitude\\
 $\Gamma$ & $2.09$ & $\pm 0.05$ & years & biennial phase-variation period\\
 $\psi$ & $2.8$ & $\pm 0.1$ & rad & biennial phase-variation phase\\
\end{tabular}
\caption{Ephemeris parameters for the times of light maxima presented in Equation 3, derived from eight years of pulse timing measurements. The fortnightly variation refers to reflex motion of the sdB from the M dwarf, while the biennial variation refers to the reflex motion of the inner sdB+dM binary due to a potential circumbinary planet.}
\end{table*}

Our best-fitting phases to the light curves serve as observed ($O$) values for the arrival times of pulsation maxima. We computed calculated ($C$) values using the linear ephemeris  
\begin{equation}
C=T_0 + PE,
\end{equation}
where $T_0$ is a reference time of light maximum, $P$ is the pulsation period at $T_0$, and $E$ is the event (pulse) number (where $E=0$ is defined as the pulse at $T_0$). We adopt as starting values for $T_0$ and $P$ the values reported in Table 1 of Barlow et al. 2011 \cite{barlow2011}. By computing the differences between $O$ and $C$, we built the O-C diagram shown in Figure 2. 
Three strong deviations from linearity are present in our data, which we model as: (i) a fortnightly sinusoidal oscillation, (ii) an upward--facing parabola, and (iii) a biennial sinusoidal phase oscillation. To quantify these structures, we fit to the O-C data the mathematical expression
\begin{align*}
&O - C = \Delta T + \Delta P E + \frac{1}{2}P\dot P E^2  \\& + A\sin\left(\frac{2\pi E}{\Pi}+\phi\right) + B\sin\left(\frac{2\pi E}{\Gamma}+\psi\right). \numbereq
\end{align*}
where $\dot{P}$ is the rate of period change; $A$, $\Pi$, and $\phi$ are the semi--amplitude, period, and phase of the fortnightly variation; and $B$, $\Gamma$, and $\psi$ are the semi--amplitude, period and phase of the biennial variation. The parameters $\Delta T$ and $\Delta P$ provide corrections to the linear ephemeris (Eq. 1) used to compute predicted $C$ values. Therefore, adjusting the previously-known parameters by these offsets allows us to construct the following updated ephemeris for the times of light maxima:
\begin{align*}
& t_{\rm max} = T_0 +  P E + \frac{1}{2}P\dot P E^2  \\& + A\sin\left(\frac{2\pi E}{\Pi}+\phi\right) + B\sin\left(\frac{2\pi E}{\Gamma}+\psi\right). \numbereq
\end{align*}
The best-fit parameters of this model are reported in Table 1. In the following subsections, we discuss the implications of each term in the model.

\subsection{Fortnightly Phase Variation}

The O-C diagram of 2011 \cite{barlow2011} was dominated by a sinusoidal variation with period 14.103 d. This oscillation was suggestive of orbital reflex motion of CS 1246 due to a low-mass stellar companion, likely an M dwarf. Follow--up radial velocity measurements with SOAR/Goodman later corroborated this result \citep{barlow2011_RV}, thereby confirming the utility of sdB pulsations for O-C studies. Our updated O-C diagram continues to show a sinusoidal oscillation due to this companion, a section of which is highlighted in the inset of Figure 2. We find an orbital period of $\Pi$ = 14.103 $\pm$ 0.003 d and semi--amplitude of $A$ = 10.7 $\pm$ 0.5 s. Our new period measurement significantly improves upon the precision of Barlow et al. 2011, and both it and the new amplitude are consistent with previous results. Therefore, we do not adjust the previous minimum companion mass estimates of $M_{\rm dM} = 0.115 \pm 0.005M_\odot$ (assuming $M_{\rm sdB} = 0.39 M_\odot$) or $M_{\rm dM} = 0.129 \pm 0.005M_\odot$ (assuming the canonical $M_{\rm sdB} = 0.47 M_\odot$) from Barlow et al. 2011 \cite{barlow2011}.

\begin{figure*}[t]
\begin{center}
\includegraphics[scale=.65]{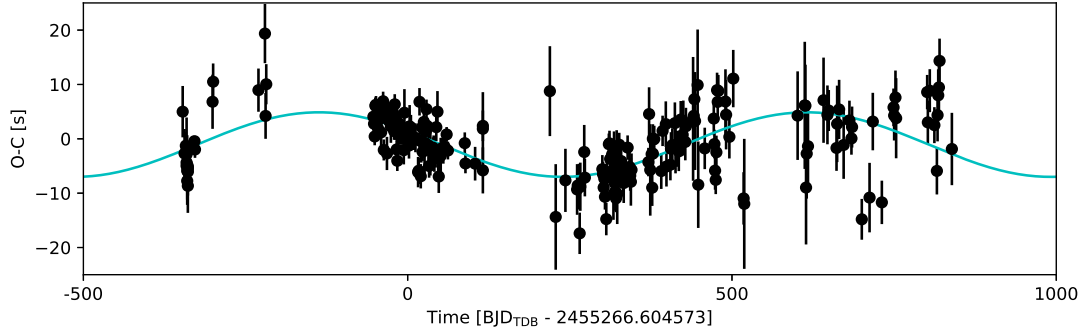}
\caption{O-C diagram with parabolic and short-period sinusoidal trends pre-whitened. The biennial oscillation occurs with a period of $2.09 \pm 0.05$ yr, consistent with reflex motion of the entire sdB+dM binary.}
\end{center}
\end{figure*}

\subsection{Biennial Phase Variation}
A second, long-period sinusoidal oscillation emerges upon adding additional timing measurements beyond 2011. Figure 3 shows the first part of the O-C diagram pre-whitened by the 14.1-d oscillation, in order to better highlight this signal. Similar to the 14.1-d variation, this biennial signal could be consistent with orbital reflex motion of the entire inner sdB+dM binary, due to the gravitational influence of a third, circumbinary object. Under this assumption, we use the binary mass function and other fundamental principles to calculate the minimum mass and orbital characteristics of the planet, which are listed in Table 2. Using the previously-measured $M_{\rm sdB} = 0.39\pm^{0.30}_{0.13}M_\odot$, we find that $m\sin i = 3.3 \pm ^{2.1}_{1.2} M_{\rm Jup}$. If, however, CS 1246 has the canonical mass $M_{\rm sdB} = 0.47M_\odot$, we find that $m\sin i = 3.7\pm{0.8}M_{\rm Jup}$. 

It is entirely possible that the observed O--C oscillation arises not from orbital reflex motion but instead from intrinsic stellar variability that is currently not understood. Traditionally, further evidence in favor of reflex motion can be found in additional pulsation modes showing the same phase variations. Unfortunately, CS 1246 shows no additional modes that could be used for this purpose. We do note that radial velocity variations of the inner sdB+dM binary would be expected if the system is orbited by a circumbinary object. Our current measurements would predict a biennial radial velocity variation with semi--amplitude of 120 $\pm$ 20 m/s. Although this signal could be measured theoretically using a high--resolution spectrograph on a large--aperture telescope, the RV signals from both the M dwarf orbit and the radial pulsations would have to be carefully removed first. This is by no means an easy task.

\subsection{Pulsation Period Change}

While the 2011 O-C diagram was consistent with a pulsational period {\em decrease}, our new data show this is no longer the case. It now appears that the apparent concave--down parabola of the former analysis was merely the local maximum of the biennial phase oscillation discussed in Section 4.2.

The new plot geometry suggests that the fundamental pulsation period of CS 1246 is {\em increasing}. Figure 4 shows the O-C diagram pre-whitened by both the 14.1-d and biennial sinusoidal oscillations, in order to better highlight the long--term variation. If such a drift is indeed due to a pulsation period change, the period would be increasing at a rate of $\dot P = (1.7 \pm 0.08) \times 10^{-12}$ s s$^{-1}$, which corresponds to 1 ms every $\sim$19 yr. { In this case, period measurements for our most recent light curves should be greater than our reported value in Table 1 by $\sim 0.4$ ms. Unfortunately, we are unable to achieve the precision required to confirm this change directly through sinusoidal fits to the light curves because (i) we do not have many recent observations and (ii) the photometric variation is quite small in the latest light curves, approaching 4 ppt. We hope to obtain several dozen SKYNET light curves in the next year to try and confirm the period change directly. 

In any case, a period} increase could be consistent with CS 1246 either (i) evolving away from the zero-age extreme horizontal branch (ZAEHB) or (ii) having recently exhausted its core He for fusion \cite{Charpinet2002}. We note that one could also interpret the parabola as an additional long--period, high--amplitude sinusoidal variation; we currently avoid such an interpretation in this work. Additional measurements in 2018 and 2019 should resolve this matter.

\begin{figure}
\vspace{-0.7cm}
\includegraphics[scale=.55]{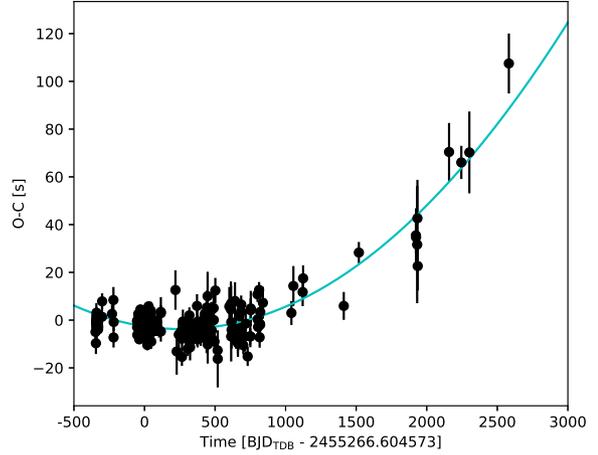}
\caption{The O-C diagram of CS 1246 with both fortnightly and biennial sinusoidal variations pre-whitened. The resulting parabola indicates a period increase of $(1.74 \pm 0.08)\times 10^{-12}$ s s$^{-1}$.}
\end{figure}

\begin{table}
\begin{tabular}{c c c c l}
 Parameter & Value & Error & Unit & Comment\\ \hline
$P$ & $2.09$ & $\pm 0.05$ & yr & orbital period\\
$K_{\rm in}$ & $120$ & $\pm 20$ & m s$^{-1}$ & RV of inner binary\\
\hline
$a$ & $1.3$ & $\pm 0.3$ & AU & orbital radius$^a$ \\
$K_{\rm out}$ & $19$ & $\pm 4$ & km s$^{-1}$ & RV of planet$^a$\\
$m\sin i$ & $3.3$ & $^{+2.1}_{-1.2}$ & $M_{\rm Jup}$ & planetary mass$^a$\\\hline
$a$ & $1.37$ & $\pm 0.02$ & AU & orbital radius$^b$ \\
$K_{\rm out}$ & $20$ & $\pm 1$ & km s$^{-1}$ & RV of planet$^b$\\
$m\sin i$ & $3.7$ & $\pm 0.8$ & $M_{\rm Jup}$ & planetary mass$^b$ \\
\end{tabular}

$^a${\footnotesize { if $M_{\rm sdB}=0.39 \, ^{+0.30}_{-0.13} \, M_\odot$ (Baade--Wesselink result).}}\\
$^b${\footnotesize { if $M_{\rm sdB}=0.47M_\odot$ (canonical mass).}}\\
\caption{Orbital parameters of the sdB$+$dM inner binary (`in') $+$ potential circumbinary planet system (`out'), as derived from pulse timing measurements.} 
\end{table}

\section{Summary}
CS 1246 is a hot subdwarf B star discovered to pulsate in 2009 through time-domain photometric and spectroscopic observations. At that time, the amplitude of its pulsations was among the largest known of all pulsating sdB stars. However, our measurements from the previous eight years indicate that this amplitude is decaying exponentially. 

CS 1246 exists in a 14.1-d binary system with an M dwarf companion, which was discovered in a 2011 pulse-timing analysis and confirmed by radial velocity measurements. Our new phase measurements show a second sinusoidal oscillation in O-C values, which is consistent with the presence of a third, low-mass, planet-sized body in the system, but we are unable to confirm this with radial velocity measurements. If this circumbinary planet truly exists, it has a minimum mass around $3-4 M_{\rm Jup}$. Additionally, the updated pulse-timing analysis falsifies the negative $\dot P$ found in 2011, instead showing an upward--facing parabolic trend consistent with an {\em increasing} period. 

As the pulsation amplitude continues to decline, noise in the O-C diagram will increase. Whether the amplitude continues to decay, approaches some asymptotic value, or eventually increases again in the future is currently unclear. In any case, we plan to continue monitoring this interesting star using the SKYNET robotic telescope network over the next few years.

\section*{Acknowledgements}

This publication has made use of the NASA Astrophysics Data System (ADS). The authors would like to thank the Student Government Association (SGA) at High Point University for providing funding to attend the Eighth Meeting on Hot Subdwarfs and Related Objects in Krak\'ow, Poland, July 2017. We would also like to thank the SOC and LOC committees, as well as the Pedagogical University of Krak\'ow, for organizing the meeting. Lastly, we are grateful for the support of High Point University and the Department of Physics.


\end{document}